\documentclass[longauth]{aa}  

\usepackage{graphicx}
\usepackage{natbib}
\bibpunct{(}{)}{;}{a}{}{,}

\usepackage{silence}
\usepackage{txfonts}
\usepackage{placeins}
\usepackage{rotating}

\usepackage[breaklinks=true]{hyperref}

\makeatletter
\renewcommand*\aa@pageof{, page \thepage{} of \pageref*{LastPage}}

\usepackage{amsmath}
\usepackage{amssymb}
\usepackage{xspace}
\usepackage{multirow}
\usepackage{tablefootnote}
\usepackage{chngcntr}
\usepackage{multicol}
\newcommand{\mum}{$\mu$m\xspace}

\newcommand{\methine}{CH\xspace}
\newcommand{\methylene}{CH$_2$\xspace}
\newcommand{\methyl}{CH$_3$\xspace}

\begin{document}

\title{The JWST Proto-PAH project: Aromatic backbones with extensive aliphatic substitution account for both the aromatic and aliphatic emission}
  
\author{N. Clark\inst{\ref{uwo}, \ref{WSpace}}, E. Peeters\inst{\ref{uwo}, \ref{WSpace}}, J. Cami\inst{\ref{uwo}, \ref{WSpace}}, G. C. Sloan\inst{\ref{stsci}, \ref{chapel}}, K. Volk\inst{\ref{stsci}}, K. E. Kraemer\inst{\ref{bostonC}}, H. L. Dinerstein\inst{\ref{UfT}}, ,M. A.  G\'{o}mez-Mu\~{n}oz \inst{\ref{barcelona1}, \ref{barcelona2}, \ref{barcelona3}}, G. M. Wahlgren\inst{\ref{stsci}}, 
N. C. Sterling\inst{\ref{georgia}}, A. Ricca\inst{\ref{Ames}, \ref{SETI}}, A. A. Zijlstra\inst{\ref{Manchester}}, J. Li\inst{\ref{Canarias1}, \ref{Canarias2}}, M. Matsuura\inst{\ref{cardiff}}, D. A. Garc\'{i}a-Hern\'{a}ndez\inst{\ref{Canarias1}, \ref{Canarias2}}, R. Sahai\inst{\ref{jpl}}, B. Aringer\inst{\ref{uppsala}, \ref{Vienna}}, C. Bhatt\inst{\ref{uwo}, \ref{WSpace}}, J. Bernard-Salas\inst{\ref{ACRI}, \ref{INCLASS}}, D. Das\inst{\ref{uwo}}, A. Manchado\inst{\ref{Canarias1}, \ref{Canarias2}, \ref{CSIC}}}

\institute{Department of Physics and Astronomy, The University of Western Ontario, London, ON N6A 3K7, Canada \label{uwo} \\
\email{nclark68@uwo.ca} \and
Institute for Earth and Space Exploration, The University of Western Ontario, London, ON N6A 3K7, Canada \label{WSpace} \and
Space Telescope Science Institute, 3700 San Martin Drive, Baltimore, MD 21218, USA\label{stsci} \and
Department of Physics and Astronomy, University of North Carolina, Chapel Hill, NC 27599-3255, USA\label{chapel}  \and
Institute for Scientific Research, Boston College, 140 Commonwealth Avenue, Chestnut Hill, MA 02467, USA\label{bostonC}  \and 
Department of Astronomy, University of Texas at Austin, Austin, TX 78712, USA \label{UfT} \and 
Departament de F\'{i}sica Qu\`{a}ntica i Astrof\'{i}sica (FQA), Universitat de Barcelona (UB), c.\ Mart\'{i} i Franqu\`{e}s, 1, 08028 Barcelona, Spain\label{barcelona1}  \and
Institut de Ci\`{e}ncies del Cosmos (ICCUB), Universitat de Barcelona (UB), c. Mart\'{i} i Franqu\`{e}s, 1, 08028 Barcelona, Spain \label{barcelona2}  \and
Institut d'Estudis Espacials de Catalunya (IEEC), Edifici RDIT, Campus UPC, 08860 Castelldefels (Barcelona), Spain \label{barcelona3}  \and
University of West Georgia, Carrollton, GA 30118, USA\label{georgia} \and 
NASA Ames Research Center, MS 245-6, Moffett Field, CA 94035-1000, USA \label{Ames} \and 
Carl Sagan Center, SETI Institute, 339 Bernardo Avenue, Suite 200, Mountain View, CA 94043, USA \label{SETI} \and 
Jodrell Bank Centre for Astrophysics, The University of Manchester, Manchester, M13 9PL, UK \label{Manchester} \and 
Instituto de Astrof\'{\i}sica de Canarias, C/ Via L\'actea s/n, E-38205 La Laguna, Spain \label{Canarias1} \and 
Departamento de Astrof\'{\i}sica, Universidad de La Laguna (ULL), E-38206 La Laguna, Spain \label{Canarias2} \and 
Cardiff Hub for Astrophysical Research and Technology (CHART), School of Physics and Astronomy, Cardiff University, The Parade, Cardiff CF24 3AA, UK \label{cardiff} \and 
Jet Propulsion Laboratory, MS 183-900, 4800 Oak Grove Dr., California Institute of Technology, Pasadena, CA 91109, USA \label{jpl} \and 
Theoretical Astrophysics, Department of Physics and Astronomy, Uppsala University, Box 516, 751 20 Uppsala, Sweden \label{uppsala} \and 
Department of Astrophysics, University of Vienna, T\"{u}rkenschanzstra{\ss}e 17, 1180 Wien, Austria \label{Vienna} \and
ACRI-ST, Centre d’Etudes et de Recherche de Grasse (CERGA), 10 Av.\ Nicolas Copernic, 06130 Grasse, France \label{ACRI} \and 
INCLASS Common Laboratory, 10 Av.\ Nicolas Copernic, 06130 Grasse, France \label{INCLASS} \and
Consejo Superior de Investigaciones Cient\'{\i}ficas (CSIC), Spain \label{CSIC}}

\titlerunning{JWST Proto-PAH project: aromatic and aliphatic emission due to single carrier population}
\authorrunning{Clark et al.}

\date{Accepted September 15th, 2026}

\abstract{We find that the aromatic and aliphatic emission bands in four carbon-rich post-asymptotic giant branch stars with class~D emission arise from a single carrier population: extensive aliphatic decoration of an aromatic skeleton, i.e.\ molecules with a coherent aromatic backbone and an unusually high aliphatic content. This conclusion is based on JWST NIRSpec and MIRI/MRS spectroscopy, covering 2.88--28.7~\mum. 

The aliphatic 3.4~\mum feature is stronger than the aromatic 3.3~\mum feature, reversing the ratio seen in sources with typical polycyclic aromatic hydrocarbon (PAH) emission. If the aromatic emission arose from nominal PAHs, the 3.3~\mum\ feature -- unavailable in previous observations -- together with the mid-infrared aromatic features, would constrain them to be large ($\gtrsim$200 carbon atoms), compact, and at least partially ionized. These properties predict a strong 8.6~\mum\ feature. However, the 8.6~\mum\ feature is weak or absent in all four sources, ruling out PAHs as the source of the aromatic emission. Instead, the emission must arise from the same carriers as the aliphatic emission. The position of the 6.9~\mum feature and the broadened 11–14~\mum out-of-plane emission independently support this. The \methylene and \methyl feature strengths vary across the sample, with the class~D2 sources showing higher \methyl content. These carriers are far more aliphatic than the PAHs of planetary nebulae and the interstellar medium. However, they retain the aromatic skeleton that produces their aromatic features. }

\maketitle
\nolinenumbers

\section{Introduction}
\label{sec:introduction}

The winds of carbon stars produce an abundance of carbonaceous species ranging from single-carbon molecules to amorphous carbon grains \citep[e.g.,][]{1987Martin, 2000Cernicharo, 2009Groenwegen}. As these stars evolve rapidly past the asymptotic giant branch (AGB) and into carbon-rich planetary nebulae (PNe), their molecular inventory changes significantly under increasingly harsh UV radiation \citep{1989VanderVeen, 1999Kwok, 2007Sloan, 2008Tielens}. Carbon stars exhibit emission from simple molecular species, whereas PNe exhibit emission from more complex molecules, such as polycyclic aromatic hydrocarbons (PAHs) and fullerenes \citep{1989Allamandola, 2000Cernicharo, 2009Groenwegen, 2010Cami}. The post-AGB phase allows us to probe this transition in detail.

Polycyclic aromatic hydrocarbon features \citep[also known as the unidentified infrared bands or UIRs; e.g.,][]{1989Allamandola} are classified according to their spectral characteristics, which provide a diagnostic of the chemical structures of their underlying carriers \citep{2002Peeters, 2004VanDiedenhoven, 2014Matsuura, 2014Sloan}. They are classified as A, B, or C, with class B being the most common in PNe. 

Several carbon-rich post-AGB stars in the Magellanic Clouds exhibit peculiar spectra inconsistent with any of the three PAH classes \citep{2014Sloan, 2014Matsuura}. In particular, the 7--9~\mum emission is exceptionally broad, and no “valley” is observed between the typical 7.7 and 8.6~\mum features. These sources are therefore classified as class D. Little is known about the nature of these carriers; however, they have been suggested to represent an intermediate stage in the chemical evolution from the carbonaceous species of AGB stars to the fully formed PAHs of PNe \citep{2014Matsuura, 2014Sloan}. Since these carriers may represent species evolving into the familiar PAHs of PNe, we refer to them as ``Proto-PAHs.''

Previous Spitzer observations were limited by the instrument's low spectral resolution and its lack of near-IR coverage. These observations, nevertheless, still made significant progress in characterizing this emission \citep{2014Sloan}. Notably, strong features at 6.9 and 7.25~\mum were identified and attributed to dust grains containing large amounts of aliphatic content \citep{2014Sloan}, and emission inconsistent with PAHs was identified in the 11--14~\mum range \citep{2014Sloan, 2014Matsuura}. The absence of near-IR coverage, however, left the aromatic component of these sources unconstrained: without the 3.3~\mum feature, the size, charge, and structure of the aromatic carriers could not be determined, and it remained unclear whether the aliphatic and aromatic emission arises from the same species or from two co-located populations. In this work we exploited the near-infrared coverage and higher spectral resolution of JWST to resolve this question. We show that the aliphatic and aromatic emission arises from a single population of carriers: extensive aliphatic decoration of an aromatic skeleton.

\section{Observations and data reduction}
\label{sec:obs_and_data_reduc}

\begin{figure*}[ht!]
\centering
\includegraphics[width=0.9\linewidth]{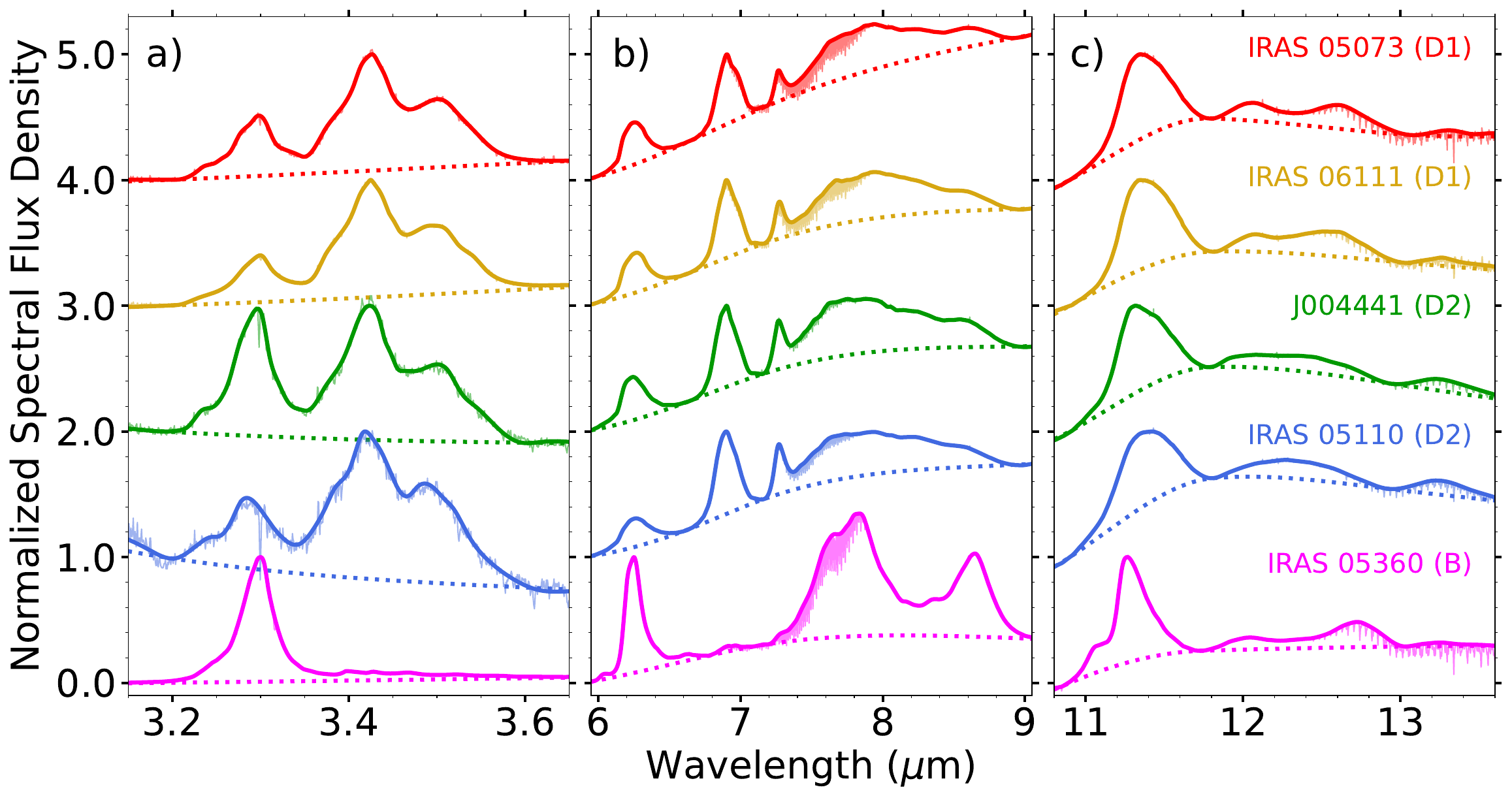}
\caption{Spectra, scaled and offset, along with the adopted continuum estimates (dotted lines). The observed spectra are shown as thin colored lines, and those with molecular features removed as thick lines of the same color. IRAS~05073 (class~D1, scale: 353.01, offset: 4) is shown in red, IRAS~06111 (class~D1, scale: 190.88, offset: 3) in orange, J004441 (class~D2, scale: 749.73, offset: 2) in green, IRAS~05110 (class~D2, scale: 880.88, offset: 1) in blue, and IRAS~05360 (class~B, scale: 25.03, offset: 0) in purple. IRAS~05360 (class~B) is included as a reference for typical PAH emission in a post-AGB or PN environment. Panel~a): Emission from the C--H stretching modes over the 3--4~\mum range. Panel~b): Emission from the C--H aliphatic bending, C--C stretching, and C--H in-plane bending modes over the 6--9~\mum range. Panel~c): Emission from the aromatic C--H OOP bending modes over the 11--15~\mum range.}
\label{fig:spectra_cont}
\end{figure*}

In this paper we examine spectroscopic observations taken with the JWST \citep{2023Gardner} NIRSpec \citep{2022Jakobsen, 2023Boker} and Mid-Infrared Instrument \citep[MIRI;][]{2023Wright}, as part of the GO3 program 4678 (PI: G. Sloan). For NIRSpec the fixed slit S200A1 (0.2\arcsec\, width), G395H grating, and F290LP filter was used, yielding 2.88--5.12~\mum spectra with a gap at 3.68--3.79~\mum at a spectral resolution of 2700). For MIRI, the data were obtained using the Medium Resolution Spectrometer (MRS) \citep{2023Argyriou}. The MRS comprises four integral field units (channels 1--4), and three grating settings (short, medium, and long, also known as A, B, C), giving 4.9--27.9~\mum spectra at a spectral resolution of 2000--3000. Further details on the observations as well as information on the data reduction process, including the application of the point fixed-pattern correction \citep[PFPC, ][]{2026Gordon}, are provided in \citet{2026Sloan}. 

The focus of this work is on the spectra of four carbon-rich post-AGB stars previously classified as exhibiting class~D PAH emission in the 6--9~\mum range \citep{2014Matsuura, 2014Sloan}: IRAS~05073--6752\footnote{short-form names of these targets are used throughout for brevity}, IRAS~06111--7023, 2MASS~J00444111--7321361, and IRAS~05110--6616. Three of the four targets are located in the Large Magellanic Cloud (LMC), while J004441 is in the Small Magellanic Cloud (SMC). Of these, IRAS~05073 and IRAS~06111 were further classified as D1 (exhibiting bands with peak wavelength near 11.3, 12.0, and 12.7~\mum), while J004441 and IRAS~05110 were classified as D2 \citep[having bands with peak wavelength near 11.4, 12.4, 13.2~\mum,][]{2014Matsuura, 2014Sloan}. Also included is IRAS~05360--7121 (in the LMC) as a reference. In contrast to the class~D features, IRAS~05360 exhibits class~B features \citep{2014Matsuura, 2014Sloan}, with strong, easily recognizable PAH features at 3.3, 6.2, 7.7, 8.6, and 11.24~\mum (Fig.~\ref{fig:spectra_cont}).

The spectra exhibit an abundance of narrow molecular emission and absorption features \citep{2026Sloan}. These were excluded to improve the clarity of our figures, treating each as unblended, with the exception of the 3.0--3.15 and 14--15~\mum ranges. In these ranges, the blended features were replaced with a cubic spline, smoothly connecting the spectrum on either side. To isolate the emission features from the underlying emission, a cubic spline was fit to the data to serve as a continuum (see Table~\ref{tab:splinewaves} for details). The fitted continua include broad plateau features common in PAH-emitting sources, extending over the 6--9 and 11--14~\mum ranges (Fig.~\ref{fig:spectra_cont}).

\begin{figure*}[ht!]
\centering
\includegraphics[width=0.9\linewidth]{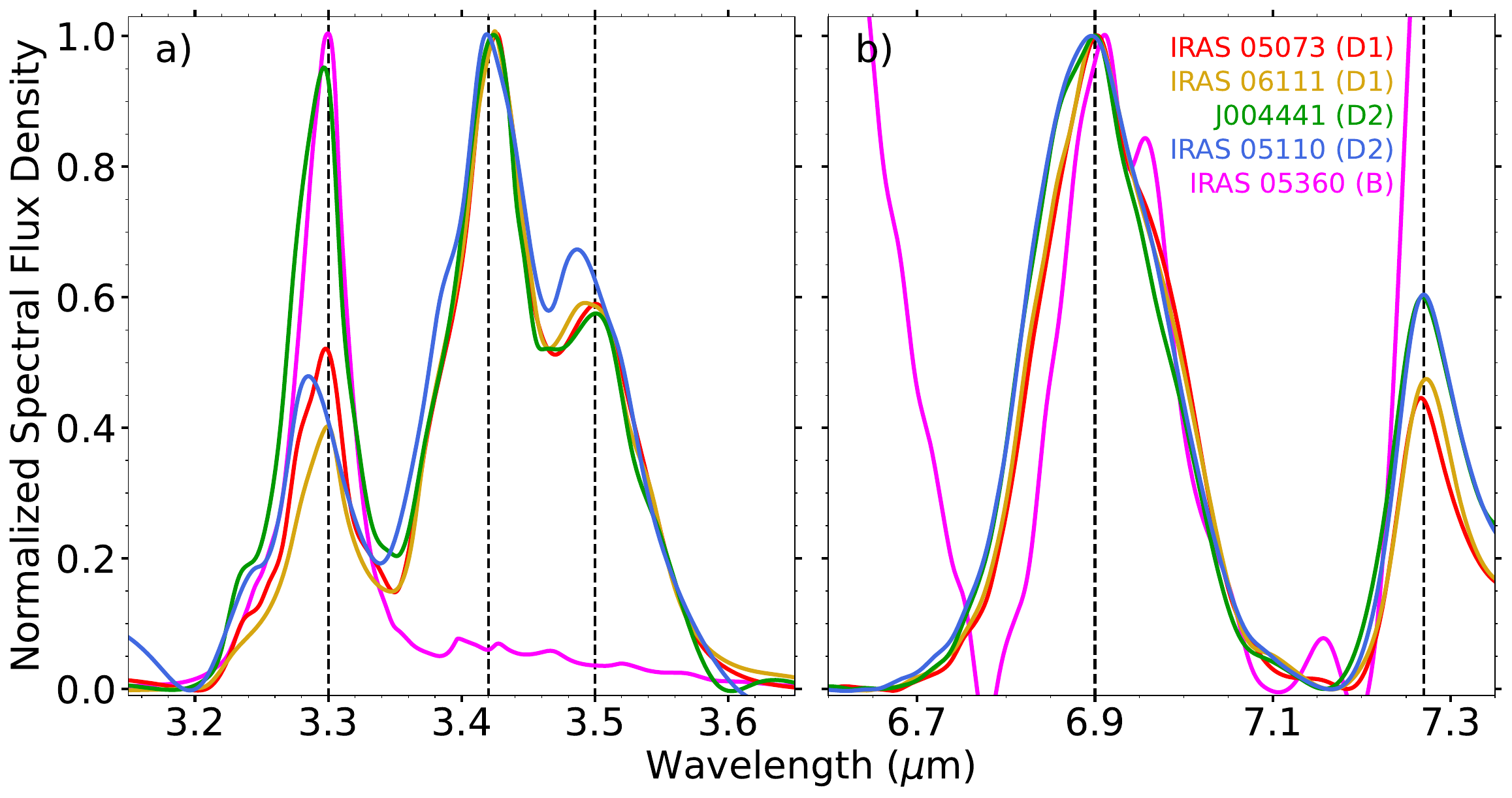}
\caption{Normalized, continuum-subtracted emission features. IRAS~05360 is included as a reference. Its aliphatic emission is much smaller than its aromatic emission. Panel~a): Emission from the C--H stretching modes over the 3.1--3.7~\mum range. This is normalized to the peak emission at 3.42~\mum, except for IRAS~05360, which is normalized to the peak emission at 3.30~\mum. The vertical dashed lines indicate the locations of the 3.3~\mum feature peak at 3.30~\mum and the main 3.4~\mum feature peaks at 3.42 and 3.50~\mum. Panel~b): Emission primarily from the C--H aliphatic bending modes over the 6.6--7.4~\mum range. This is normalized to the peak emission at 6.90~\mum. The aromatic 7.7~\mum feature is also partially included, blending with the red wing of the 7.25~\mum feature and dominating it in the case of IRAS~05360. This spectrum also includes the end of a feature at 6.7~\mum. The vertical dashed lines indicate the locations of the 6.9 and 7.25~\mum feature peaks at 6.90 and 7.27~\mum.}
\label{fig:features}
\end{figure*}

\section{Results}
\label{sec:Results}

Figure~\ref{fig:spectra_cont} illustrates the spectral differences between class~D and class~B sources \citep{2014Matsuura, 2014Sloan}. One key distinction is the prominence of aliphatic emission in class~D spectra, with strong features at 3.4, 6.9, and 7.25~\mum. 

Such aliphatic emission is typically a fraction of the strength of the aromatic emission and has previously been associated with aliphatic functional groups attached to larger hydrocarbon structures rather than with free carbon chains and super-hydrogenated PAHs \citep{1996Bernstein, 1996Joblin, 2000Chiar, 2008Pino, 2012Carpentier, 2013Jones}. This is examined in Sect.~\ref{subsec:singlepopulation}. This section focuses on these three features as well as the 3.3~\mum feature.

\subsection{Spectral characteristics}

Figure~\ref{fig:features} shows the normalized, continuum-subtracted 3.3, 3.4, 6.9, and 7.25~\mum features of the class~D spectra. The profiles are overall very similar, with most differences arising from the relative strengths of individual feature subcomponents.

Figure~\ref{fig:features}a shows the 3.3 and 3.4~\mum features. As the spectrum of IRAS~05360 demonstrates in Fig.~\ref{fig:features}a, the 3.3~\mum feature is typically much stronger than the 3.4~\mum feature. In the class~D spectra, however, this is reversed: the 3.4~\mum feature is consistently much stronger than the 3.3~\mum feature. Overall, the 3.4~\mum feature profile is similar across the class~D sources. The class~D2 source, IRAS~05110, exhibits some subtle deviations: its 3.52~\mum peak is enhanced relative to the 3.42~\mum peak, with both located at slightly bluer wavelengths, and the more pronounced shoulder at 3.38~\mum produces a slightly broader blue rise overall. Further variation is seen in the peak position of the 3.3~\mum feature, ranging from 3.28~\mum for the class~D2 source, IRAS~05110 -- consistent with the blueshift seen in its 3.4~\mum components -- to 3.30~\mum for the other sources.

\begin{table*}
\begin{center}
\caption{Integrated intensity ratios for key emission features and components. The intensities are listed in Table~\ref{tab:intensity}.}
\label{tab:ratios}
\begin{tabular}{ccccc}
\hline
Feature Ratio & IRAS~05073 (D1) & IRAS~06111 (D1) & J004441 (D2) & IRAS~05110 (D2) \\
\hline

3.3/11.2 & 
0.053 $\pm$ 0.003 & 
0.047 $\pm$ 0.002 & 
0.085 $\pm$ 0.005 & 
0.023 $\pm$ 0.003 \\ 

6.2/11.2 & 
0.63 $\pm$ 0.02 & 
0.70 $\pm$ 0.02 & 
0.68 $\pm$ 0.02 & 
0.56 $\pm$ 0.01 \\ 

3.38/3.42 &
0.41 $\pm$ 0.02 &
0.44 $\pm$ 0.01 &
0.50 $\pm$ 0.03 &
0.66 $\pm$ 0.05 \\

7.25/6.9 aliph. & 
0.20 $\pm$ 0.05 & 
0.21 $\pm$ 0.04 & 
0.28 $\pm$ 0.01 & 
0.28 $\pm$ 0.01 \\ 

3.4 aliph./$\Sigma$(3.3, 3.4) & 
0.77 $\pm$ 0.03 & 
0.81 $\pm$ 0.02 & 
0.66 $\pm$ 0.04 & 
0.77 $\pm$ 0.08 \\ 

6.9 aliph./6.9 & 
0.82 $\pm$ 0.04 & 
0.83 $\pm$ 0.03 & 
0.91 $\pm$ 0.03 & 
0.91 $\pm$ 0.02 \\ 
\hline
\end{tabular}
\end{center}
\end{table*}

The 6.9 and 7.25~\mum features show similar spectral characteristics across the sources (Fig.~\ref{fig:features}b). Notably, when normalized to the 6.9~\mum feature strength, the 7.25~\mum features are of similar strength within the class~D2 sources and likewise within the class~D1 sources. However, the class~D1 features are overall weaker than those of the class~D2 sources. These and other key ratios are listed in Table~\ref{tab:ratios}.

\begin{figure*}[ht!]
\centering
\includegraphics[width=0.9\linewidth]{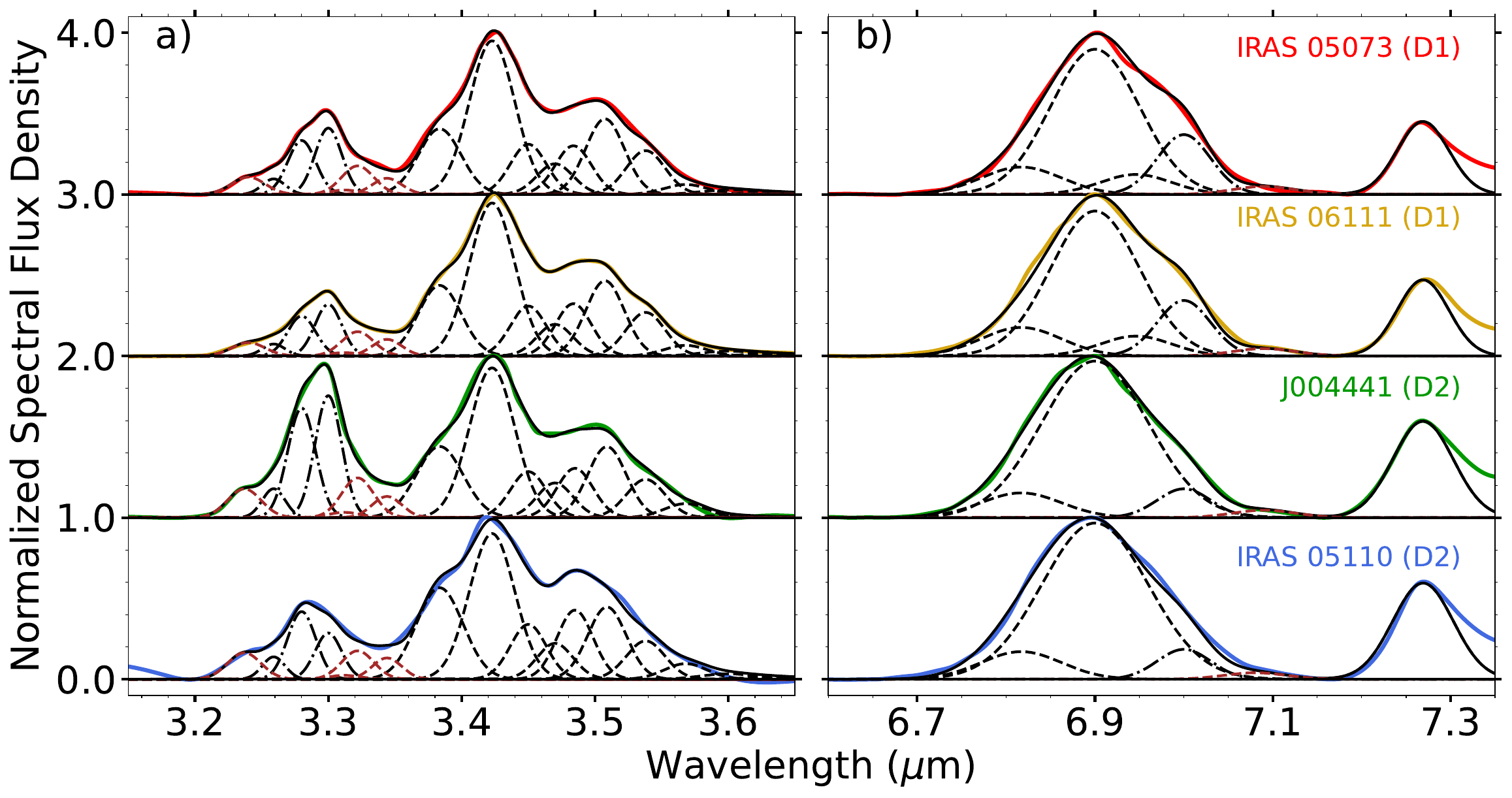}
\caption{Decomposition of normalized, continuum-subtracted emission features, with an offset added for clarity. The spectra are normalized to the peak emission at 3.42~\mum (panel~a) and at 6.90~\mum (panel~b). IRAS~05073 (class~D1; scale: (a) 383.07, (b) 35.22; offset: 3) is shown in red, IRAS~06111 (class~D1; scale: (a) 206.09, (b) 16.36; offset: 2) in orange, J004441 (class~D2; scale: (a) 703.80, (b) 50.98; offset: 1) in green, and IRAS~05110 (class~D2; scale: (a) 754.13, (b) 20.35; offset: 0) in blue. Gaussians corresponding to aromatic vibrations are shown with dash-dotted black lines, those for aliphatic vibrations are shown with dashed black lines, and those for olefinic vibrations are shown with dashed brown lines. The sum of the fitted Gaussians is shown with a solid black line.}
\label{fig:components}
\end{figure*}

\subsection{Assignments and decomposition}

Carbonaceous emission features can be broadly classified according to the type of chemical bonding of the C and H responsible for them. Aromatic C--H bonds occur on sp$^2$-hybridized carbon in conjugated ring structures, as in PAHs. Aliphatic C--H bonds occur on sp$^3$-hybridized carbon, either in noncyclic chain structures, such as methyl (\methyl) and methylene (\methylene) groups, or are incorporated into a ring system, such as a saturated CH$_2$ group within an otherwise aromatic polycyclic skeleton (e.g., indene, phenalene, and 1H-cyclopent[cd]indene), or as an aromatic ring carbon that has been superhydrogenated. Olefinic C--H bonds, also sp$^2$-hybridized, occur on nonaromatic carbon chains containing C=C double bonds. 
Vibrational assignments for these modes are well established \citep[][]{1963Silverstein, 1980Socrates, 1989Allamandola, 1998Ristein, 2007Dartois}, and these are adopted throughout. A specific reference is given only where a particular assignment is at issue. Aromatic, aliphatic, and olefinic C--H vibrational modes often overlap in wavelength. The 3.3~\mum feature is primarily associated with aromatic C--H stretching, with some contribution from olefinic C--H stretching, whereas the 3.4~\mum feature arises from aliphatic C--H stretching \citep{1986JourdaindeMuizon, 1992Geballe, 1994Geballe}. The 6.9 and 7.25~\mum features arise primarily from aliphatic C--H bending modes, with some weak aromatic and olefinic contributions between them. These include asymmetric bending, such as scissoring modes near 6.9~\mum, and symmetric bending near 7.25~\mum.

\begin{table}
\begin{center}
\caption{Fitting parameters of the Gaussian decompositions. Each component is assigned a vibrational bond type -- olefinic (Olef.), aromatic (Arom.), or aliphatic (Aliph.) -- along with a more specific vibrational assignment. Vibrational assignments are abbreviated as follows: stretching (str.), bending (bnd.), symmetric (sym.), and asymmetric (asym.). All parameters are expressed in \mum. Any variation in fitting parameters across sources is noted.}
\label{tab:gaussian}
\begin{tabular}{ll}
\hline
Peak Position & Vibrational \\
(Standard Deviation) & Assignment \\
\hline
\multicolumn{2}{c}{\textbf{3.3, 3.4~\mum features}} \\
3.240 (0.0125) $^1$ & Olef. \methylene asym. str.  \\
3.259 (0.0085) & Arom. \methine str.  \\
3.280 (0.0105) & Arom. \methine str.  \\
3.300 (0.0100) & Arom. \methine str.  \\
3.312 (0.0120) & Olef. \methine str.  \\
3.322 (0.0120) & Olef. \methine str.  \\
3.344 (0.0110) & Olef. \methylene sym. str.  \\
3.3835 (0.0160) $^2$& Aliph. \methyl asym. str. \\
3.4228 (0.0170) & Aliph. \methylene asym. str.  \\
3.450 (0.0135) & Aliph. \methine str.  \\
3.470 (0.0135) & Aliph. \methine str.  \\
3.4838 (0.0135) $^3$& Aliph. \methyl sym. str.  \\
3.5075 (0.0145) $^4$& Aliph. \methylene sym. str. \\
3.538 (0.0145) & Aliph. \methylene sym. str.  \\
3.568 (0.0160) & Aliph. str. red wing  \\
3.600 (0.0200) & Aliph. str. red wing  \\
3.640 (0.0240) & Aliph. str. red wing  \\
\hline
\multicolumn{2}{c}{\textbf{6.9, 7.25~\mum features}} \\
6.817 (0.045) & Aliph. \methylene, \methyl asym. bnd.  \\
6.900 (0.050) $^5$& Aliph. \methylene, \methyl asym. bnd. \\
6.946 (0.040) & Aliph. \methylene, \methyl asym. bnd.  \\
7.000 (0.0300) & Arom. \methine bnd.  \\
7.090 (0.0350) $^6$& Olef. \methylene bnd. \\
7.269 (0.0290) $^7$& Aliph. \methyl sym. bnd. \\
\hline\\[-10pt]
\end{tabular}
$^1$~Class~D2: peak at 3.237\\
$^2$~Class~D2: std. dev. of 0.0175\\
$^3$~Class~D2: peak at 3.4850\\
$^4$~Class~D2; peak at 3.5085\\
$^5$~Class~D2; std. dev. of 0.051\\
$^6$~IRAS~05110: peak at 7.080\\
$^7$~Class~D2: std. dev. of 0.0295\\
\end{center}\end{table}

To aid our analysis, the 3.3, 3.4, 6.9, and 7.25~\mum features were decomposed using a series of Gaussians, each corresponding to a known C--H vibrational assignment. These decompositions are shown in Fig.~\ref{fig:components}. The fits were calculated with a linear least squares approach. To reduce degeneracy, the Gaussian components were incrementally fitted for each feature: amplitudes, widths, and peak positions were first determined for regions well represented by a single Gaussian component, with additional blended components subsequently fitted. Care was also taken to ensure that the components reported as weak or narrow in the literature remained so, with minimal variation in peak position and width of each component across the different sources. Table~\ref{tab:gaussian} lists the fitted positions, standard deviations, and vibrational modes. The integrated intensities of the components, derived from the fitting, are listed in Table~\ref{tab:amplitudes}. This work aims to isolate the aliphatic emission from the aromatic and olefinic contributions and to resolve the subcomponents that constrain the CH$_3$/CH$_2$ balance and the proximity of aliphatic groups to aromatic rings. Every component does not require a unique assignment. 

Gaussians corresponding to aliphatic vibrations fit the 3.4~\mum feature well, while combinations of aromatic and olefinic vibrations fit the 3.3~\mum band well. Of note, variation in the relative strengths of the components at 3.38, and beyond 3.50~\mum, recreate the observed 3.4~\mum profile variations. Gaussians corresponding to aliphatic vibrations fit both the 6.9 and 7.25~\mum features, although subtle variations in the 6.9~\mum band arise from non-aliphatic emission. The 7.25~\mum feature has a broad red wing, which we attribute to blending with the strong aromatic 7.7~\mum feature \citep{2002Peeters}.

\section{Discussion}
\label{sec:discussion}

\subsection{A single population: Aliphatic groups on an aromatic skeleton}
\label{subsec:singlepopulation}

The strong aliphatic features in these sources could arise in two ways. Either the aliphatic and aromatic emission come from two independent populations -- PAHs carrying few aliphatic side-groups, as is typical of astronomical PAHs \citep[e.g.,][]{2016Yang, 2017Yang}, alongside a separate, predominantly aliphatic component (see Sect.~\ref{subsec:structure_of_carriers} for a possible example) -- or they come from a single population of molecules bearing both aromatic and aliphatic C--H bonds. For the remainder of this paper, we refer to PAHs carrying few aliphatic side-groups, as is typical of astronomical PAHs, as "nominal PAHs." We show below that the observations require the second option: no distribution of nominal PAHs can account for the observed aromatic emission, while both the observed aromatic and aliphatic emission are consistent with a single carrier population.

First, we considered the population of nominal PAHs needed to produce the observed aromatic features and find that no such distribution of nominal PAHs. The size, charge, and edge structure were each considered in turn. The ratio of the 3.3 and 11.2~\mum bands (hereafter 3.3/11.2 ratio) traces the size of PAHs \citep[e.g.,][]{1993Schutte, 2012Ricca, 2016Croiset, 2021Knight}. Using the diagnostics involving the 3.3 and 11.2~\mum features from \citet{2020Maragkoudakis}, adjusted for the anharmonic correction of \citet{2023Lemmens}, the observed ratios (Table~\ref{tab:ratios}) indicate that the PAHs contain at least 200 carbon atoms. The relative strengths of the 6.2 and 7.7~\mum features probe the PAH charge \citep{1984Leger, 1989Allamandola, 2008Galliano}. The 6.2~\mum feature is strong relative to the 11.2~\mum feature (6.2/11.2 = 0.56--0.70; see Table~\ref{tab:ratios}), indicating that the PAHs are at least partially ionized \citep{2020Maragkoudakis}. The 7.7~\mum feature is blended with the red wing of the 7.25~\mum aliphatic feature and contaminated by molecular absorption, so we did not attempt to measure it. Its strength, however, is evident in Fig.~\ref{fig:spectra_cont}. The relative strength of the 11--14~\mum features probes the edge structure of PAHs \citep{2001Hony, 2025Khan}. Since the 11.2~\mum feature is strong relative to the 12.7 and 13.5~\mum features (Fig.~\ref{fig:spectra_cont}), these PAHs are compact, consisting mostly of straight edges. Large, compact, ionized PAHs are expected to produce a strong 8.6~\mum PAH feature \citep{2008Bauschlicher, 2009Bauschlicher, 2012Ricca}. Yet, the 8.6~\mum feature is weak or absent in all four sources. The aromatic features therefore cannot be produced by any distribution of nominal PAHs: the very properties required to explain the 3.3, 6.2, 7.7, and 11.2~\mum emission predict a strong 8.6~\mum feature that is weak or absent.

We next considered the 6.9~\mum feature, attributed to aliphatic C--H asymmetric bending, specifically a C--H scissoring mode \citep{1963Silverstein, 1980Socrates, 1989Allamandola, 1998Ristein, 2007Dartois}. The position of this feature depends on the chemical structure surrounding the aliphatic C--H bond undergoing the scissoring motion --- specifically, whether it occurs on an isolated aliphatic chain or adjacent to an aromatic ring. For C--H scissoring modes on aliphatic chains, the feature is located at 6.82~\mum. However, when the aliphatic C--H scissoring mode occurs adjacent to an aromatic ring, the feature is redshifted to 6.9~\mum \citep{1990Colthup, 2013Sandford, 2017Materese}. The decomposition of the 6.9~\mum feature in each object shows that the majority of the emission is centered at 6.9~\mum, with only a small contribution from emission located at 6.82~\mum. We thus conclude that the majority of the aliphatics responsible for the 6.9~\mum feature are adjacent to aromatic rings.

Lastly, we considered the emission in the 11--14~\mum range. For PAHs consisting only of aromatic bonds, this emission arises from C--H out-of-plane (OOP) bending modes \citep{1984Leger, 1989Allamandola, 1999Hudgins, 2001Hony}. There is no expected emission in this wavelength range from aliphatics or olefinics \citep{1990Colthup, 2013Sandford, 2020Dartois}. Thus, under the two-population hypothesis, all distinct emission features in the 11--14~\mum range would be attributed to the PAHs. Molecules with aromatic rings bearing C--H bonds and aliphatic substituents on other rings still produce OOP features. However, the addition of aliphatics breaks the vibrational coupling of the OOP modes, resulting in emission across the 11--14~\mum range that becomes increasingly spread out and blended \citep{1990Colthup, 2013Sandford, 2020Dartois}. This is precisely what is observed in the class~D spectra (Fig.~\ref{fig:spectra_cont}). For example, the 11.2~\mum PAH feature, which typically peaks between 11.20 and 11.25~\mum, is here shifted to 11.33~\mum and has a much broader red wing than is typical. Additionally, emission is present at 12.4 and 13.2~\mum, which has not been attributed to PAH OOP modes \citep{2001Hony, 2014Matsuura, 2025Khan} and has previously been observed in spectra with strong aliphatic 6.9 and 7.25~\mum features \citep{2014Sloan, 2014Matsuura}. This broadening and blending of the OOP features in the class~D spectra therefore also supports the single-population scenario.

\subsection{Spectral variation among the sources}

While Sect.~\ref{subsec:singlepopulation} establishes that a single carrier population produces both the aromatic and aliphatic emission in these sources, the class D sample nonetheless shows real source-to-source spectral diversity, which we examine here. We attribute the variation in aliphatic emission among the class~D sources to differences in the type of aliphatic content present. As an example, aliphatic \methyl stretching modes produce emission at 3.38~\mum, while \methylene stretching modes produce emission at 3.42~\mum. Similarly, the 7.25~\mum feature is primarily attributed to \methyl bending modes, whereas both \methylene and \methyl bending modes contribute to 6.9~\mum emission \citep{1990Colthup, 1998Ristein, 2007Dartois}. The class~D1 sources have smaller 3.38/3.42~\mum component ratios than the class~D2 sources, and the 7.25/6.9~\mum ratio is also larger in the class~D2 sources. Both lines of evidence suggest that the class~D2 sources have relatively higher \methyl content. 

Some variation is also seen among the sources themselves. The 3.3/11.2 ratio is markedly higher in J004441 (0.085) than in the other three sources (0.023–0.053; \ref{tab:ratios}), which correspondingly drives its notably lower 3.4 aliph./$\Sigma$(3.3,3.4) ratio (0.66 vs. 0.77–0.81 elsewhere). J004441 is also the only source in our sample located in the SMC rather than the LMC. The fractional contribution of the 3.3~\mum feature to the total PAH luminosity is well established to increase with decreasing metallicity, attributed to inhibited PAH growth, which suppresses the population of larger grains at low metallicities while leaving the smallest, 3.3~\mum-emitting carriers comparatively unaffected \citep{Whitcomb2024, Whitcomb2026, Lai2025, Tarantino2025}. Critically, this dependence is steep even between the LMC (Z$\sim$0.4-0.5Z$\odot$) and SMC (Z$\sim$0.2Z$\odot$), despite both being subsolar. This offers a plausible explanation for J004441's outlying 3.3/11.2 ratio without requiring a distinct aromatic carrier population. Consistent with this picture, J004441's 6.2/11.2 ratio (0.68), which traces predominantly PAH charge rather than size, falls squarely within the range spanned by the LMC sources (0.56–0.70), as expected if the metallicity effect is specific to the size-sensitive 3.3~\mum feature rather than a wholesale difference in this source's carrier population. We caution, however, that an equivalent metallicity dependence has not yet been established for the aliphatic features. If the 3.4, 6.9, and 7.25~\mum features are similarly sensitive to host metallicity, the source-to-source variation discussed above may be influenced by both metallicity and differences in aliphatic content rather than reflecting aliphatic content alone.

\subsection{The structure of the proto-PAH carriers}
\label{subsec:structure_of_carriers}

Large molecules or dust grains with mixed aromatic and aliphatic content have been proposed as carriers of the unidentified infrared emission (UIRs) in place of PAHs \citep[see, e.g.,][]{1981Duley, 2011Kwok}. We briefly discuss these alternatives before turning to the implications for our sources.

Carbonaceous dust grains are excellent candidate carriers to explain the aliphatic absorption features seen along some lines of sight \citep{1991Sandford, 1994Pendleton, 2002Pendleton, 2007Dartois, 2013Chiar}. However, they are less suited to explain the UIR emission characteristics. The unidentified infrared emission points to carriers of much smaller heat capacity, which can be stochastically heated to high temperatures by individual photons, rather than reaching the lower equilibrium temperatures of bulk grains \citep[e.g.,][]{1984Sellgren, 2001DraineLi, 2008Tielens}. 

The emission spectra provide additional constraints on the molecular structure of the carriers. In the astronomical context, a PAH molecule is a coherent aromatic skeleton, potentially containing heteroatom substitutions (e.g., nitrogen) in the ring, bearing aliphatic substituents such as methyl or methylene side groups, or with super-hydrogenated edges or surfaces \citep[a ``dirty'' PAH; see, e.g.,][]{1989Allamandola, 1999Hudgins, 2008Tielens}. Such a species has a discrete vibrational spectrum and a low heat capacity. Under stochastic heating it emits the aromatic features at their characteristic positions, with aliphatic features appearing as satellites \citep[see, e.g.,][]{1999Allamandola, 2010Bauschlicher, 2014Boersma, 2014Rosenberg, 2020Mattioda}. A mixed-aromatic organic nanoparticle (MAON), by contrast, is an internally bulk-structured nanoparticle: a 3D amorphous network of small aromatic units cross-linked by aliphatic chains, in which the structure is dominated by the aliphatic linkages rather than by a coherent aromatic backbone \citep{2011Kwok}. Mixed-aromatic organic nanoparticles are also compositionally more heterogeneous than PAHs, potentially incorporating heteroatoms such as O, N, and S within the network in addition to C and H \citep{2011Kwok}. However, with respect to the emission characteristics that are the focus of this work, the key distinction between the two families is structural: the coherent, planar aromatic backbone of a PAH versus the disordered, aliphatically cross-linked network of a MAON. The corresponding difference in heat capacity and vibrational mode structure govern how each emits rather than any difference in elemental composition.

The sharpness and fixed positions of the 6.2, 7.7, and 11.2~\mum features require a coherent aromatic skeleton, and the low aliphatic fractions inferred for the general UIR population \citep[$\sim$2--9\%;][]{2012Li, 2016Yang, 2017Yang} place these carriers firmly on the (dirty-)PAH side of this divide. Experimental and theoretical PAH spectra compare to the observed UIRs \citep[see, e.g.,][]{1999Allamandola, 2010Bauschlicher,   2020Mattioda} well, further supporting this interpretation.

Because a MAON is an aliphatic-dominated network, its emission spectrum is necessarily dominated by the aliphatic features rather than by the aromatic-dominated feature set that defines the UIRs. To the best of our knowledge, no spectrum of a MAON has yet been published. The mixed-carrier models draw their plausibility from bulk hydrogenated amorphous carbon (HAC) and quenched carbonaceous composite (QCC) laboratory analogues, without spectroscopically modeling the proposed particle itself. We therefore did not adopt the mixed-carrier hypothesis for the general UIR features. 

The four sources we studied, however, are precisely where this distinction becomes interesting. They are at an aliphatic-rich, transitional phase of post-AGB evolution, in which aliphatic features weaken relative to the aromatic ones as sources evolve toward the PN stage \citep{1996Joblin, 1999Kwok}. Their aliphatic content -- evident in the reversed 3.4/3.3 ratio and the exceptional strength of the 6.9~\mum feature -- is far higher than in general UIR sources, placing them, among known carriers, closest to the aliphatic-rich regime the mixed-carrier models describe. Yet, they retain sharp aromatic features at 3.3, 6.2, 7.7, and 11.2~$\mu$m -- the signature of an intact aromatic skeleton. The weakness or absence of the 8.6~\mum feature, far from contradicting this, is precisely what distinguishes these carriers from nominal PAHs. A critical open question is whether these carriers approach the MAON-like structural regime or remain aromatic-skeleton molecules carrying an unusually heavy aliphatic decoration. A quantitative determination of their aliphatic fraction and the structural questions it raises will be the subject of a forthcoming paper.

\section{Conclusions}
\label{sec:conclusions}

We investigated JWST NIRSpec and MIRI/MRS spectroscopy of four carbon-rich post-AGB stars with class~D emission, covering the 3-–4~\mum region for the first time in these objects. We find that the aliphatic and aromatic emission arise from a single population of carriers: extensive aliphatic decoration of an aromatic skeleton, i.e.,\ molecules with a coherent aromatic backbone and an unusually high aliphatic content.

The 3.4~\mum aliphatic feature is stronger than the neighboring 3.3~\mum aromatic feature, a reversal of the ratio seen in typical PAH sources, and the 6.9 and 7.25~\mum features likewise dominate their surroundings. We decomposed the primarily aliphatic 3.4, 6.9, and 7.25~\mum features and the primarily aromatic 3.3~\mum feature into components consistent with established aromatic, aliphatic, and olefinic C--H vibrational assignments.

The aromatic features require carriers that are large, compact, and at least partially ionized. Such PAHs should produce a strong 8.6~\mum feature. Its weakness or absence in all four sources rules out any distribution of nominal PAHs as the source of the aromatic emission in these targets, which must therefore arise from the same carriers as the aliphatic emission. This conclusion is supported by the position of the 6.9~\mum feature, indicative of aliphatic groups adjacent to aromatic rings, and by the 11--14~\mum emission, consistent with aromatic C--H OOP bending modes strongly perturbed by nearby aliphatic structure.

We identify variation in the aliphatic emission between the class~D1 and class~D2 sources, consistent with the variation in the relative \methyl/\methylene content and, therefore, in the degree of methyl substitution among the carriers in these sources.

These carriers are far more aliphatic than the PAHs seen in PNe and the interstellar medium. However, they retain the aromatic skeleton that produces their aromatic features. Further observations of additional proto-PAH candidate sources, along with complementary laboratory and theoretical spectroscopy, will help clarify the origin and fate of this aliphatic-rich material.

\begin{acknowledgement}
    
This work is based on observations made with the NASA/ESA/CSA James Webb Space Telescope. All of the data presented in this article were obtained from the Mikulski Archive for Space Telescopes (MAST) at the Space Telescope Science Institute. These observations are associated with program \#4678 (DOI: 10.17909/zs5n-2t21).
N.C., E.P., J.C., C.B., and D.D. acknowledge support from the Canadian Space Agency (CSA, 24JWGO3B01), and the Natural Sciences and Engineering Research Council of Canada. 
Support for US investigators (G.C.S., H.L.D., K.E.K., R.S., N.S., and G.M.W.) in program \#4678 was provided by NASA through a grant from the Space Telescope Science Institute, which is operated by the Association of Universities for Research in Astronomy, Inc., under NASA contract NAS 5-03127. 
J. L., D.A.G.H. and A.M. acknowledge support from the State Research Agency (AEI) of the Spanish Ministry of Science, Innovation, and Universities (MICIU) of the Government of Spain, and the European Regional Development fund (ERDF), under grant PID2023-147325NB-I00/AEI/10.13039/501100011033. This publication is based upon work from COST Action CA21126 - Carbon molecular nanostructures in space (NanoSpace), supported by COST (European Cooperation in Science and Technology).
M.A.G.M. acknowledges being funded by the European Union (ERC, CET-3PO, 101042610). Views and opinions expressed are, however, those of the author(s) only and do not necessarily reflect those of the European Union or the European Research Council Executive Agency. Neither the European Union nor the granting authority can be held responsible for them. M.A.G.M. also acknowledges financial support from grant CEX2024-001451-M funded by MICIU/AEI/10.13039/501100011033.
R.S.’s contribution to the research described here was carried out at the Jet Propulsion Laboratory, California Institute of Technology, under a contract with NASA (80NM0018D0004), and partially funded by award  JWST-GO-04678 from the STScI under NASA contract NAS5-03127.
\end{acknowledgement}

\bibliographystyle{aa}
\bibliography{bib}

\begin{appendix}

\onecolumn

\begin{multicols}{2}

\section{Additional analysis value tables}

Table~\ref{tab:splinewaves} lists the wavelengths for the anchor points used in the continuum determination. Tables~\ref{tab:amplitudes} and \ref{tab:intensity} list the integrated intensity of the Gaussian components used in the decomposition and of the features used in the ratios of Table~\ref{tab:ratios}.
\bigskip\bigskip\bigskip\bigskip
\end{multicols}

\begin{table*}[h]
\begin{center}
\caption{Wavelengths of the anchor points used for the continuum spline fitting for each spectra, in units of ~\mum.}
\label{tab:splinewaves}
\begin{tabular}{ccccc}
\hline
IRAS~05073 (D1) & IRAS~06111 (D1) & J004441 (D2) & IRAS~05110 (D2) & IRAS~05360 (B) \\
\hline
2.9066 & 2.9066 & 2.9066 & 2.9066 & 2.9066 \\
3.1915 & 3.1915 & 3.1915 & 3.1915 & 3.1015 \\
3.6654 & 3.6854 & 3.6654 & 3.6854 & 3.6854 \\
3.8837 & 3.8837 & 3.8837 & 3.8837 & 3.9000 \\
4.2851 & 4.2851 & 4.2851 & 4.2851 & 4.2480 \\
4.6000 & 4.6000 & 4.6000 & 4.6000 & 4.6000 \\
4.9568 & 4.9568 & 4.9568 & 4.9000 & 4.9568 \\
\hline
5.1196 & 5.1196 & 5.1196 & 5.1196 & 5.1196 \\
5.4795 & 5.4795 & 5.4771 & 5.4771 & 5.4681 \\
5.9431 & 5.9431 & 5.9431 & 5.9000 & 5.9678 \\
6.6550 & 6.6550 & 6.6550 & 6.6550 & 6.7900 \\
7.1823 & 7.1600 & 7.1600 & 7.1600 & 7.0950 \\
8.9163 & 8.9222 & 8.9435 & 8.9134 & 9.1580 \\
8.9800 & 8.9700 & 9.0150 & 8.9800 & 9.2500 \\
9.0400 & 9.0373 & 9.0597 & 9.0420 & 9.3521 \\
9.2200 & 9.2200 & 9.1500 & 9.2200 & 9.4380 \\
\hline
9.5750 & 9.5750 & 9.5980 & 9.5743 & 9.4390 \\
9.9430 & 9.9764 & 9.9781 & 9.9430 & 9.9430 \\
10.2390 & 10.2390 & 10.2230 & 10.2030 & 10.2220 \\
10.8090 & 10.8950 & 10.7720 & 10.7940 & 10.8326 \\
10.8810 & 10.9348 & 10.9260 & 10.8920 & 10.8682 \\
11.7750 & 11.7942 & 11.7729 & 11.8049 & 11.7290 \\
11.8080 & 11.8371 & 11.7960 & 11.8086 & 11.7474 \\
13.0450 & 12.9960 & 12.9186 & 12.9140 & 13.0498 \\
13.9900 & 13.7490 & 13.9733 & 13.8620 & 13.9352 \\
\hline
\end{tabular}
\end{center}
\end{table*}

\begin{table*}[h]\begin{center}
\caption{Integrated intensity of the Gaussian components of each fitted feature, in units of $1\times10^{-17}~W/m^2$.}
\label{tab:amplitudes}
\begin{tabular}{cccccc}
\hline
Peak Position & Vibrational Assignment & IRAS~05073 (D1) & IRAS~06111 (D1) & J004441 (D2) & IRAS~05110 (D2) \\
\hline
\multicolumn{6}{c}{\textbf{3.3, 3.4~\mum features}} \\
3.240 & Olef. \methylene asym. str. & 0.25 $\pm$ 0.01 & 0.36 $\pm$ 0.01 & 0.23 $\pm$ 0.01 & 0.20 $\pm$ 0.02 \\
3.259 & Arom. \methine str. & 0.15 $\pm$ 0.01 & 0.22 $\pm$ 0.02 & 0.15 $\pm$ 0.01 & 0.11 $\pm$ 0.01 \\
3.280 & Arom. \methine str. & 0.64 $\pm$ 0.02 & 0.88 $\pm$ 0.02 & 0.71 $\pm$ 0.01 & 0.41 $\pm$ 0.03 \\
3.300 & Arom. \methine str. & 0.74 $\pm$ 0.02 & 1.07 $\pm$ 0.02 & 0.74 $\pm$ 0.02 & 0.26 $\pm$ 0.04 \\
3.312 & Olef. \methine str. & 0.06 $\pm$ 0.01 & 0.08 $\pm$ 0.01 & 0.04 $\pm$ 0.01 & 0.03 $\pm$ 0.01 \\
3.322 & Olef. \methine str. & 0.38 $\pm$ 0.01 & 0.60 $\pm$ 0.01 & 0.29 $\pm$ 0.01 & 0.19 $\pm$ 0.01 \\
3.344 & Olef. \methylene sym. str. & 0.19 $\pm$ 0.03 & 0.37 $\pm$ 0.02 & 0.14 $\pm$ 0.02 & 0.13 $\pm$ 0.04 \\
3.3835 & Aliph. \methyl asym. str. & 1.11 $\pm$ 0.05 & 2.23 $\pm$ 0.06 & 0.72 $\pm$ 0.04 & 0.86 $\pm$ 0.06 \\
3.4228 & Aliph. \methylene asym. str. & 2.70 $\pm$ 0.03 & 5.01 $\pm$ 0.05 & 1.44 $\pm$ 0.04 & 1.30 $\pm$ 0.05 \\
3.450 & Aliph. \methine str. & 0.69 $\pm$ 0.01 & 1.28 $\pm$ 0.02 & 0.34 $\pm$ 0.02 & 0.38 $\pm$ 0.02 \\
3.470  & Aliph. \methine str. & 0.42 $\pm$ 0.01 & 0.80 $\pm$ 0.02 & 0.26 $\pm$ 0.01 & 0.25 $\pm$ 0.01 \\
3.4838 & Aliph. \methyl sym. str. & 0.66 $\pm$ 0.02 & 1.31 $\pm$ 0.02 & 0.36 $\pm$ 0.01 & 0.47 $\pm$ 0.03 \\
3.5075 & Aliph. \methylene sym. str. & 1.08 $\pm$ 0.01 & 2.00 $\pm$ 0.02 & 0.55 $\pm$ 0.02 & 0.52 $\pm$ 0.02 \\
3.538 & Aliph. \methylene sym. str. & 0.62 $\pm$ 0.03 & 1.14 $\pm$ 0.03 & 0.29 $\pm$ 0.02 & 0.27 $\pm$ 0.04 \\
3.568 & Aliph. str. red wing & 0.15 $\pm$ 0.02 & 0.30 $\pm$ 0.03 & 0.12 $\pm$ 0.02 & 0.12 $\pm$ 0.03 \\
3.600 & Aliph. str. red wing & 0.09 $\pm$ 0.02 & 0.17 $\pm$ 0.03 & 0.02 $\pm$ 0.03 & 0.05 $\pm$ 0.03 \\
3.640 & Aliph. str. red wing & 0.05 $\pm$ 0.02 & 0.10 $\pm$ 0.04 & 0.01 $\pm$ 0.02 & 0.03 $\pm$ 0.04 \\
\hline
\multicolumn{6}{c}{\textbf{6.9, 7.25~\mum features}} \\
6.817 & Aliph. \methylene, \methyl asym. bnd. & 3.5 $\pm$ 0.2 & 7.9 $\pm$ 0.3 & 2.2 $\pm$ 0.1 & 6.1133 $\pm$ 0.1 \\
6.900 & Aliph. \methylene, \methyl asym. bnd. & 20.1 $\pm$ 0.2 & 43.3 $\pm$ 0.8 & 18.0 $\pm$ 0.2 & 45.0 $\pm$ 0.2 \\
6.946 & Aliph. \methylene, \methyl asym. bnd. & 2.2 $\pm$ 0.1 & 4.7 $\pm$ 0.2 & 0.0 $\pm$ 0.1 & 0.0 $\pm$ 0.1 \\
7.000 & Arom. \methine bnd. & 4.8 $\pm$ 0.4 & 9.7 $\pm$ 0.6 & 1.6 $\pm$ 0.1 & 4.2 $\pm$ 0.3 \\
7.090 & Olef. \methylene bnd. & 0.7 $\pm$ 0.3 & 1.5 $\pm$ 0.4 & 0.5 $\pm$ 0.1 & 1.0 $\pm$ 0.2 \\
7.269 & Aliph. \methyl sym. bnd. & 5.3 $\pm$ 1.2 & 11.9 $\pm$ 2.3 & 5.7 $\pm$ 0.3 & 14.1 $\pm$ 0.7 \\
\hline
\end{tabular}
\end{center}\end{table*}

\begin{table*}[h]
\begin{center}
\caption{Integrated intensities of the features and components appearing in the ratios of Table~\ref{tab:ratios}, in units of $1\times10^{-17}~W/m^2$.}
\label{tab:intensity}
\begin{tabular}{ccccc}
\hline
Feature & IRAS~05073 (D1) & IRAS~06111 (D1) & J004441 (D2) & IRAS~05110 (D2) \\
\hline

3.3 & 
2.4 $\pm$ 0.1 &
3.6 $\pm$ 0.1 &
2.3 $\pm$ 0.1 &
1.3 $\pm$ 0.2 \\

3.38 & 
1.11 $\pm$ 0.05 &
2.23 $\pm$ 0.06 &
0.72 $\pm$ 0.04 &
0.86 $\pm$ 0.06 \\

3.42 & 
2.70 $\pm$ 0.03 &
5.01 $\pm$ 0.05 &
1.44 $\pm$ 0.04 &
1.30 $\pm$ 0.05 \\

3.4 & 
8.2 $\pm$ 0.2 &
15.5 $\pm$ 0.3 &
4.4 $\pm$ 0.2 &
4.5 $\pm$ 0.3 \\

$\Sigma$(3.3, 3.4) &
10.6 $\pm$ 0.3 &
19.1 $\pm$ 0.4 &
6.7 $\pm$ 0.3 &
5.9 $\pm$ 0.5 \\

6.2 & 
28.4 $\pm$ 0.8 &
53.0 $\pm$ 1.0 &
18.3 $\pm$ 0.4 &
32.4 $\pm$ 0.6 \\

6.9 aliph & 
25.8 $\pm$ 0.5 &
55.8 $\pm$ 1.3 &
20.2 $\pm$ 0.4 &
51.1 $\pm$ 0.4 \\

6.9 & 
31.4 $\pm$ 1.3 &
67.0 $\pm$ 2.3 &
22.3 $\pm$ 0.6 &
56.3 $\pm$ 1.0 \\

7.25 & 
5.3 $\pm$ 1.2 &
11.9 $\pm$ 2.3 &
5.7 $\pm$ 0.3 &
14.1 $\pm$ 0.7 \\

11.2 & 
45.2 $\pm$ 0.9 &
75.7 $\pm$ 1.2 &
26.9 $\pm$ 0.4 &
58.1 $\pm$ 0.8 \\
\hline
\end{tabular}
\end{center}
\end{table*}

\end{appendix}

\end{document}